 \documentclass[aps,prb,reprint,groupedaddress]{revtex4-1}
\usepackage[dvipdfmx]{graphicx}% Include figure files
\usepackage{dcolumn}% Align table columns on decimal point
\usepackage{bm}% bold math
\usepackage{soul}

\begin{document}

% Use the \preprint command to place your local institutional report
% number in the upper righthand corner of the title page in preprint mode.
% Multiple \preprint commands are allowed.
% Use the 'preprintnumbers' class option to override journal defaults
% to display numbers if necessary
%\preprint{}

%Title of paper
\title{Nuclear resonant scattering of synchrotron radiation by physisorbed Kr on TiO$_{2}$(110) surfaces in multilayer and monolayer regimes}

% repeat the \author .. \affiliation  etc. as needed
% \email, \thanks, \homepage, \altaffiliation all apply to the current
% author. Explanatory text should go in the []'s, actual e-mail
% address or url should go in the {}'s for \email and \homepage.
% Please use the appropriate macro foreach each type of information

% \affiliation command applies to all authors since the last
% \affiliation command. The \affiliation command should follow the
% other information
% \affiliation can be followed by \email, \homepage, \thanks as well.
\author{Akihiko Ikeda}\email[Current address: Institute for Solid State Physics, University of Tokyo, Kashiwa, Chiba, Japan, email: ]{ikeda@issp.u-tokyo.ac.jp}
%\affiliation{Institute for Solid State Physics, The University of Tokyo, Kashiwa, Chiba, Japan}
\affiliation{Institute of Industrial Science, The University of Tokyo, Komaba, Tokyo, Japan}

\author{Taizo Kawauchi}
\affiliation{Institute of Industrial Science, The University of Tokyo, Komaba, Tokyo, Japan}

\author{Masuaki Matsumoto}
\affiliation{Natural Science Division, Tokyo Gakugei University, Koganei, Tokyo, Japan}
\affiliation{Institute of Industrial Science, The University of Tokyo, Komaba, Tokyo, Japan}

\author{Tatsuo Okano}
\affiliation{Institute of Industrial Science, The University of Tokyo, Komaba, Tokyo, Japan}

\author{Katsuyuki Fukutani}\email[Corresponding author: ]{fukutani@iis.u-tokyo.ac.jp}
\affiliation{Institute of Industrial Science, The University of Tokyo, Komaba, Tokyo, Japan}

\author{Xiao Wei Zhang}
\affiliation{Institute of Material Structure Science, High Energy Accelerator Research Organization (KEK), Tsukuba, Ibaraki, Japan}
%\affiliation{High Energy Accelerator Research Organization (KEK), Tsukuba, Ibaraki, Japan}

%\affiliation{IMSS, KEK, Tsukuba, Ibaraki, Japan}

\author{Yoshitaka Yoda}
\affiliation{Japan Synchrotron Radiation Research Institute (JASRI)/SPring-8, Koto, Hyogo, Japan}
%\affiliation{JASRI/SPring-8, Koto, Hyogo, Japan}

%
%\homepage[]{Your web page}
%\thanks{}
%\altaffiliation{}

%Collaboration name if desired (requires use of superscriptaddress
%option in \documentclass). \noaffiliation is required (may also be
%used with the \author command).
%\collaboration can be followed by \email, \homepage, \thanks as well.
%\collaboration{}
%\noaffiliation

\date{\today}

\begin{abstract}
% insert abstract here
Physisorbed Kr layers on TiO$_{2}$(110) surfaces were investigated by means of nuclear resonant scattering (NRS) of synchrotron radiation at Kr thicknesses ranging from multilayer to monolayer.  The NRS intensity was measured as a function of the Kr exposure, from which the NRS signal corresponding to monolayer  was estimated as 0.23 cps.  The time spectra measured at various thicknesses showed a monotonous decay without any quantum beat features.  The recoiless fraction $f$ evaluated from the analysis of the time spectrum revealed a substantial reduction upon temperature rise from 19 to 25 K.  As its origin, an order-disorder phase transition of the monolayer Kr is proposed.

\end{abstract}

% insert suggested PACS numbers in braces on next line
% insert suggested keywords - APS authors don't need to do this
%\keywords{}

%\maketitle must follow title, authors, abstract, \pacs, and \keywords
\maketitle

% body of paper here - Use proper section commands
% References should be done using the \cite, \ref, and \label commands
\section{Introduction}

Since the discovery of the M\"ossbauer effect, the M\"ossbauer spectroscopy (MS) has been extensively employed as a probe to the local magnetism, electronic structures and the phase of the matter through the hyperfine fields at the resonant nuclei and recoilless fractions $f$. Nuclear resonant scattering (NRS) is an extension of MS in the time domain utilizing the synchrotron radiation (SR), which made possible a variety of spectroscopic studies as well as x-ray diffraction. \cite{Rohlsberger} By virtue of the brilliance and the energy tunability of SR, NRS has not only opened the accessibility to the resonant nuclei such as $^{83}$Kr,\cite{Baron} $^{129}$Xe,\cite{Klobes} $^{197}$Au,\cite{Kishimoto} $^{151}$Eu,\cite{Matsuoka} and so forth, but also extended its target to nanostructures, thin films, \cite{Kawauchi1, Slezak1, Slezak2} and even to monolayer materials, although this is still limited to Fe/W(110).\cite{RohlsbergerPRL, Sladecek, Vogl2007, Partykajankowska, Vogl2009}

As a two-dimensional material, a physisorbed monolayer system is also fascinating due to its unique phase transition. Kr on exfoliated graphite was extensively studied for its commensurate-incommensurate (CI) phase transition and order-disorder (OD) phase transition, which is an experimental realization of the three-state Potts model.\cite{Kardar} Physisorbed atoms can also be used as a nondestructive local probe of the electronic and magnetic properties at surfaces, for instance, through the techniques such as the photoelectron spectroscopy of adsorbed Xe \cite{Wandelt, Getzlaff, Guo} or the nuclear magnetic resonance of adsorbed Cd.\cite{Bellini, Cottenier} In recent years, furthermore, the van der Waals system is extensively studied with experiments \cite{Ikeda1} and \textit{ab initio} calculations \cite{Bagus, Silva1, Chen} for their importance in the organic optoelectric devices. Nevertheless, due to the difficulties of probing the monolayer materials with x rays, studies on monolayer Kr were so far limited to Kr on exfoliated graphite. \cite{Johnson} The recent improvement of the brilliance and emittance of SR has made it possible to explore a Kr monolayer system on single-crystal surfaces with NRS. We anticipate that Kr on TiO$_{2}$(110) should show a simple phase transition by analogy to the OD phase transition of Si(001). \cite{Tabata, Matsumoto} We also expect that a large electric-field gradient (EFG) on the TiO$_{2}$(110) surface may lead to a hyperfine splitting of nuclear levels of $^{83}$Kr. \cite{Bellini, Cottenier}

In this paper, we investigated physisorbed Kr layers on TiO$_{2}$(110)  by means of x-ray reflectivity (XR) and NRS. We obtained appreciable NRS intensity even in the monolayer regime. We also obtained a series of time spectra of NRS by physisorbed Kr layers as a function of Kr exposure. We evaluated the recoilless fraction, which shows a large difference between 19 and 25 K in contrast to the case of solid Kr and monolayer Kr on the exfoliated graphite surface. \cite{Kolk, Johnson} We discuss the OD phase transition as a possible origin of the observed temperature dependence of the NRS intensity.

\begin{figure}[b]
\begin{center}
\includegraphics[scale=0.4]{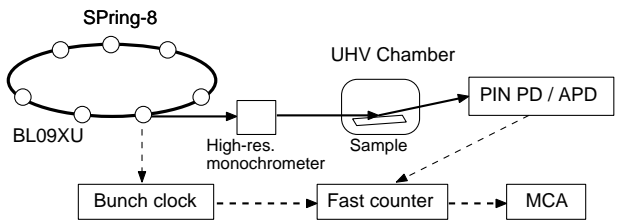} 
\caption{Schematic of the experimental setup of NRS measurements from physisorbed layers in an ultrahigh vacuum chamber. \label{exp}}
\end{center}
\end{figure}

\section{Experiment}

An ultrahigh vacuum (UHV) chamber with a base pressure of $2\times10^{-8}$ Pa was installed in the beamline BL09XU, SPring-8, Japan. The whole UHV chamber was mounted on mechanical linear guides in the horizontal and vertical directions, both of which were perpendicular to the SR beam for the positioning of the sample.  With an aid of a UHV type radiation reflectomator, \cite{Kawauchi2} the sample was rotated with respect to the SR beam with a precision of $\pm0.01^{\circ}$ for the $\theta$-$2\theta$ configuration. The SR beam entered and escaped from the UHV chamber via Be windows.

A rutile TiO$_2$(110) single crystal (Shinko-sha, $14 \times 7 \times 0.5$ mm$^3$) held with a Ta plate was attached to a Cu block fixed at the head of a closed-cycle He compression type refrigerator. The temperature of the sample was monitored with a Chromel-Alumel type thermocouple spot welded to the Ta plate that holds the sample substrate. The substrate was annealed at 500 K in UHV until LEED revealed a sharp 1x1 pattern.  Since the sample was still transparent and not light blue in color, the sample was not substantially reduced with this treatment.  The sample surface was also characterized with ultraviolet photoemssion spectroscopy in a separate chamber, which showed no defect-related in-gap state indicating that little oxygen vacancies were produced on the surface. \cite{Fukada, Diebold}  The sample was then transferred through the air to the UHV chamber at the beamline in Spring-8, where the sample was annealed in UHV before the Kr exposure and NRS experiments. The physisorbed layers of Kr were formed on the TiO$_{2}$(110) surface by backfilling the chamber with the Kr gas via the variable leak valve. We used the Kr gas with natural abundance and the isotope-enriched Kr gas, which contain the  $^{83}$Kr isotope of 11 \% and 75 \%, respectively.

The SR with an energy of 9.4 keV was incident  on the sample surface with $\theta$ of 0 to 1.0$^{\circ}$ in the $[\bar{1}10]$ direction after it was monochromized with a high-resolution monochromator \cite{Zhang} to an energy width of about 30 meV around the resonance energy. The SR excites the nuclear state of $^{83}$Kr from the ground state of $I=9/2$  with the $(+)$ parity to an excited state of $I=7/2$ with the $(+)$ parity of which the natural lifetime is 212 ns. \cite{Ruby} The reflected SR was detected outside the UHV chamber either with a PIN photodiode (PD) or eight channel avalanche photodiodes (APD) for the XR measurements or the NRS measurements, respectively. The schematic of the experimental setup is drawn in Fig. \ref{exp}.

The SR was in the several bunches mode (D mode), in which we used 12 bunches of 1.6 mA each for the NRS measurements. The time interval between the bunches was 342.1 ns. The APD and the fast electronics were utilized for the time resolved detection of the prompt (nonresonant) and delayed (nuclear resonant) signals of the reflected SR with a time resolution of less than 0.3 ns. \cite{Kawauchi1} The time spectra of the NRS were measured by the accumulation of the delayed signal with a multichannel analyzer. The time spectra of the NRS were measured by repeated accumulations of the delayed signal for 40 min with a multichannel analyzer. The energy of SR was adjusted to the resonance before each accumulation, and the NRS intensity was confirmed to be stable between the accumulations within the statistical uncertainty, which guarantees the stability of the sample during the accumulations. The time spectra for the samples of 20 L at 19 K, 6 L at 19 K, and 6 L at 25 K were measured for total acquisition times of 11, 17 and 16 hours, respectively, and the data are shown as sums over total accumulations in the present paper. We also confirmed the reproducibility of the sample preparation and the time spectrum data for the sample of 6 L at 19 K during another SR beam time.  The deduced effective thickness was reproducible within 10 \%.

\begin{figure}[t]
\begin{center}
\includegraphics[scale=0.55]{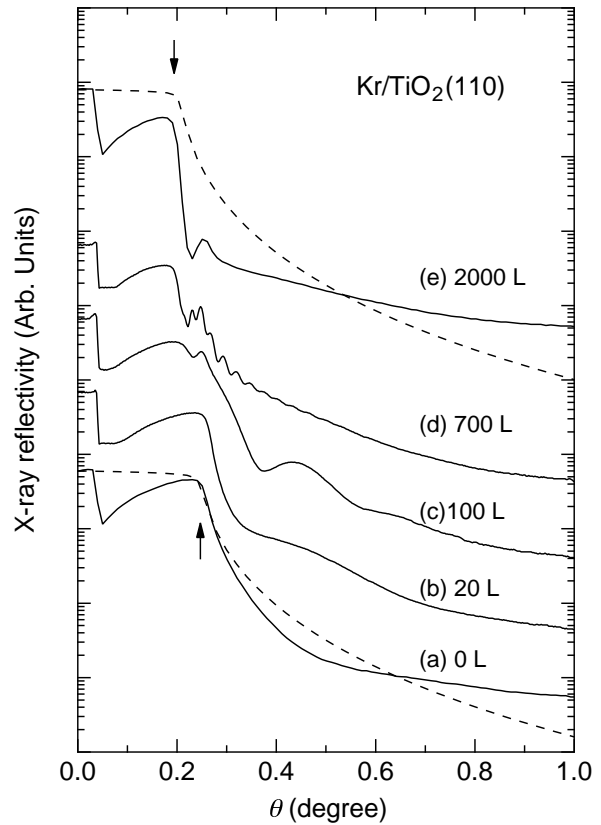} 
\caption{XR from physisorbed Kr layers as a function of $\theta$ at Kr exposure of 0 to 2000 L. Dashed curves are the calculated XR for the flat surfaces of rutil TiO$_{2}$ and solid Kr. \label{xrc}}
\end{center}
\end{figure}

\begin{figure}[t]
\begin{center}
\includegraphics[scale=0.55]{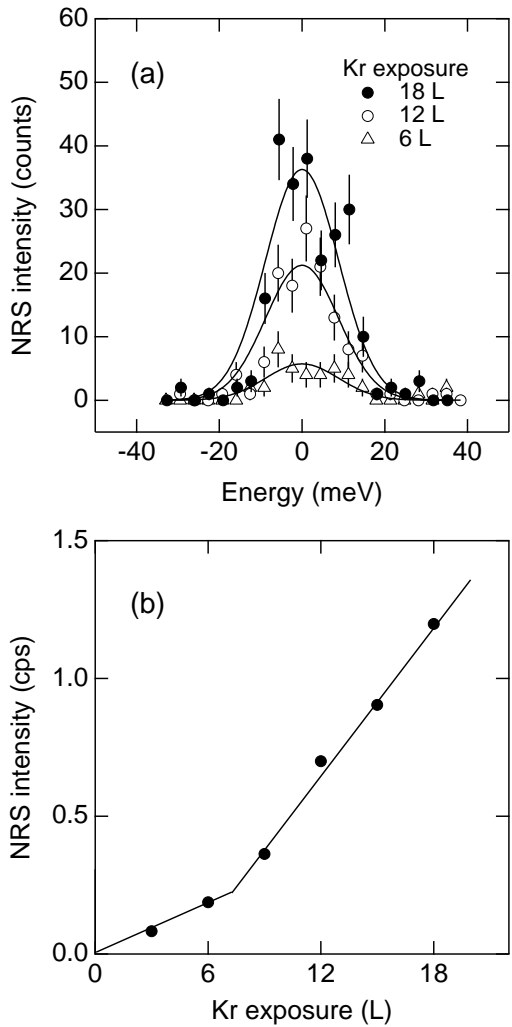}
\caption{(a) Nuclear resonant spectra at 9.4 keV by $^{83}$Kr nuclei in Kr physisorbed layers at Kr gas exposures of 6, 12, 18 L. The $^{83}$Kr isotope enriched (75 \%) Kr gas was used. Each points was obtained with the accumulation for 30 sec. (b) The deduced NRS intensity as a function of the exposure of the isotope enriched Kr gas.\label{delay}}
\end{center}
\end{figure}

\begin{figure}[t]
\begin{center}
\includegraphics[scale=0.55]{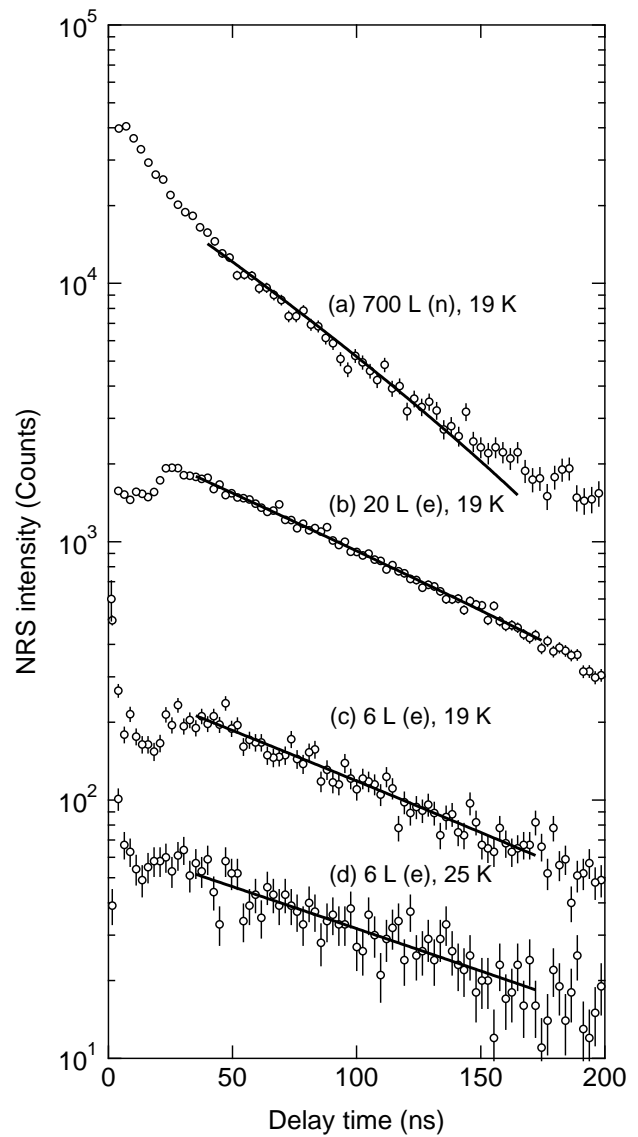} 
\caption{Time spectra of NRS of a 9.4 keV x ray from $^{83}$Kr in physisorbed layers of Kr on a TiO$_{2}$(110) surface (open circles). Solid curves are fits with Eq. (\ref{Bessel}). The spectrum (a) is multiplied by 20 for a clarity. (n) and (e) indicate that the natural or isotope enriched Kr gas was used, respectively.\label{ts}}
\end{center}
\end{figure}

\section{Results}
To confirm the formation of the Kr multilayers on the TiO$_{2}$(110) surface at 19 K \textit{in situ}, a series of XR curves were obtained as a function of the Kr exposure (1 L = $1.33\times10^{-4}$ Pa$\cdot$s), where we used the Kr gas with natural abundance. The results are shown in Fig. \ref{xrc} along with the simulated XR curves. \cite{Henke} For the XR of the clean TiO$_{2}$(110) the critical angle $\theta_{\mathrm{C}}$ was observed to be 0.257$^{\circ}$, which coincides with the calculated value of 0.262$^{\circ}$ for the flat surface of rutile TiO$_{2}$ ($\rho=4.27$ g/cm$^{3}$) as indicated by the arrow pointing to the solid and dashed curves in Fig. \ref{xrc}(a). After the exposure of Kr exceeding 2000 L as shown in Fig. \ref{xrc}(e), the observed $\theta_{\mathrm{C}}$ was 0.209$^{\circ}$, which also agrees well with the calculated value of 0.200$^{\circ}$ for the flat surface of Kr ($\rho=3.40$ g/cm$^{3}$). It is also indicated by simulated results that $\theta_{\mathrm{C}}$ is $\sim0.26^{\circ}$ at the Kr thickness of 10 nm and that it becomes $0.20^{\circ}$ at above 50 nm. The Kr film of 1 nm grows with a Kr exposure of 10 L, assuming that the sticking probability is 1.0 \cite{Head, SchlichtingSS} and considering the impinging rate $\phi=p/\sqrt{2\pi mkT}$, where $p$, $m$, $k$ and $T$ are gas pressure, mass of a Kr atom, the Boltzmann constant, and the temperature of the exposed Kr gas (= Temperature of the UHV chamber). In Figs. \ref{xrc}(b)-\ref{xrc}(d), the transition of $\theta_{\mathrm{C}}$ from $\sim0.26^{\circ}$ to $\sim0.20^{\circ}$ is observed with increasing Kr exposure from 20 L to 700 L, in good agreement with the simulated results. By these results, the formation of the Kr layer on the TiO$_{2}$(110) surface at 19 K was confirmed \textit{in situ}.

Subsequently, the delayed signal integrated over 30 ns to 340 ns of the reflected SR from Kr on TiO$_{2}$(110) was obtained by varying the energy of the incident SR beam around 9.4 keV as shown in Fig \ref{delay}(a). The spectra were measured at $\theta=0.25^{\circ}$ for various exposures of the isotope-enriched Kr gas, where the intensity of the prompt signal was $1.7\times10^{7}$ cps.  Intensity maxima appeared at around 9.4 keV. The shape of the spectra represents the energy distribution of the SR beam monochromized by the high resolution monochromator in good agreement with the previous study. \cite{Zhang} The spectra were tentatively fitted with a Gaussian function, $I(E)=A_{0}\exp\{-(E/a)^{2}\}$, where $E$ denotes the SR energy. With a width parameter fixed at $a=12.5$ meV, we obtained $A_{0}$ as the intensity of NRS at the center of the resonance as a function of the Kr exposure, which is plotted in Fig. \ref{delay}(b).

As shown in Fig. \ref{delay}(b), at above 7.5 L, the NRS intensity as a function of the exposure of Kr gas shows a linear dependence on the Kr exposure with a slope of 0.091 cps/L with some intercept at a Kr exposure of 0 L. We analyze the results in terms of the coverage dependence of the condensation probability of Kr. Here we assume that the NRS intensity is proportional to the total amount of the physisorbed Kr, which is validated by previous studies \cite{Kolk, Johnson} as is mentioned later. The density of the closed-packed monolayer of Kr is $7.2\times10^{18}$ atoms/m$^{2}$.  Since the condensation probability of Kr beyond the first layer is 1.0, \cite{Head, SchlichtingSS} this Kr density corresponds to the Kr exposure of 3.4 L calculated with the impinging rate. Therefore, we estimate the effective NRS intensity per a close-packed Kr monolayer to be $\sim0.29$ cps. In Fig. \ref{delay} (b), we read the value of 0.225 cps  at 7.5 L, which we regard as the completion of the first layer of Kr. On TiO$_{2}$(110), Kr may well commensurately physisorb according to the lattice spacing of the substrate, of which density is estimated to be 0.74 times as high as the density of the close-packed monolayer of Kr. The NRS intensity from the commensurate monolayer of Kr on the TiO$_{2}$(110) surface is then estimated to be 0.23 cps being consistent with the observed value of 0.225 cps at 7.5 L. From the slope at $0\sim7$ L in Fig. \ref{delay}(b), we deduce the initial sticking probability $s_{\mathrm{i}}$ to be $\sim0.3$, which is reasonably comparable to the reported value of 0.22 for $s_{\mathrm{i}}$ of Kr on Ru(001). \cite{SchlichtingJCP}

\begin{table*}[!t]
\caption{Experimental conditions and the results of the analysis of the NRS time spectra from the physisorbed Kr layers. Portion of $^{83}$Kr in Kr gas $\rho$, substrate temperature $T$, incident angle of the SR beam $\theta$, intensity of the prompt signal $I_{\mathrm{P}}$, intensity of the delayed signal$I_{\mathrm{D}}$, and effective thickness deduced from the time spectra with Eq. (\ref{Bessel}) $\chi$.}
\label{tab}
\begin{ruledtabular}
\begin{tabular}{llllllll}
 Exposure (L) & $\rho$ (\%) & $T$ (K) & $\theta$ (degree)  & $I_{\mathrm{P}}$ (10$^{8}$ cps)  & $I_{\mathrm{D}}$ (cps) & $I_{\mathrm{D}}/I_{\mathrm{P}}$ (10$^{-9}$)&$\chi$ (-)\\
 \hline

700 & 11.5 (n) & 19 & 0.21 & 0.8  & 1.1 & 13.8 & $2.22\pm0.07$ \\ 
20 & 75 (e) & 19 & 0.22 & 1.9  & 2.9 & 15.3 & $1.10\pm0.02$ \\ 
6 & 75 (e) & 19 & 0.25 & 1.86 & 0.24 & 1.30 & $0.88\pm0.07$ \\ 
6 & 75 (e) & 25 & 0.25 & 0.9 & 0.055 & 0.61 & $0.54\pm0.11$ \\ 

\end{tabular}
\end{ruledtabular}
\end{table*}

A series of time spectra of NRS by Kr layers as a function of the Kr exposure were obtained as shown in Fig. \ref{ts}. The time spectrum shown in Fig. \ref{ts} (a) was obtained for the Kr layer formed with the exposure of the Kr gas with natural abundance for 700 L at the substrate temperature of 19 K. The time spectra shown in Fig. \ref{ts}(b) and \ref{ts}(c) were obtained for the Kr layers formed with the exposure of the isotope enriched Kr gas for 20 L and 6 L at 19 K, respectively. The time spectrum in Fig. \ref{ts} (d) was obtained at a substrate temperature of 25 K, after the Kr layer was formed with the exposure of the isotope-enriched Kr gas for 6 L at 19 K. The incident angle $\theta$ and the intensity of the prompt signal $I_{\mathrm{p}}$ were optimized for every experimental condition so that the experimental count rates are maximized, which are summarized in Table \ref{tab}. The NRS intensity at the time range from 0 to $\sim30$ ns in Fig. \ref{ts} were depleted for the APD was insensitive in this time period not recovering from the detection of the intense prompt signal at 0 ns. Therefore, the data points in this region were not used in the analysis.

As a rule of thumb, the slopes of the time spectra in Fig. \ref{ts} increase with increasing Kr exposure, indicating that the speed up effect due to the dynamical effect is in charge. On the other hand, no appreciable quantum beat structure was found. We speculated several origins of the absence of the quantum beats, although we have not specified which is the main cause, \cite{dt_ikeda} namely, (i) the insufficient strength of EFG applied on $^{83}$Kr nuclei on the surface, (ii) inhomogeneous EFG applied on $^{83}$Kr nuclei due to the variation of adsorption site due to, for example, incommensurate adsorption, (iii) EFG with lower symmetry applied on $^{83}$Kr nuclei allowing the more complex transitions in $^{83}$Kr nuclei. \cite{Partykajankowska, Grunsteudel}

We analyzed the obtained time spectra in terms of the dynamical effect with the form \cite{Johnson}
\begin{equation}
I(\tau, \chi)=I_{0}(\chi/\tau)\exp\left(-\tau\right)\left|J_{1}\left(\sqrt{4\chi \tau}\right)\right|^{2},
\label{Bessel}
\end{equation}
where $\tau$ is time in the unit of the natural lifetime of the excited state of $^{83}$Kr. \cite{Kolk} The effective thickness is expressed as $\chi=n\sigma_{0}f/4$, where $n$ and $\sigma_{0}$ are the number of the resonant nuclei/cm$^{2}$ and the absorption cross section at resonance. The time spectra in Fig. \ref{ts}(b)-\ref{ts}(d) were well fitted with Eq. (\ref{Bessel}). The deviation of the time spectrum in Fig. \ref{ts} (a) from Eq. (\ref{Bessel}) may be a consequence of the variation of $\chi$ due to the noticeable roughness of the thick Kr layer formed with the Kr exposure of 700 L, as is also suggested by the XR result in Fig. \ref{xrc}(d).

The deduced $\chi$, the intensity of the delayed signals (the NRS intensity) $I_{\mathrm{D}}$, and the ratio $I_{\mathrm{D}}/I_{\mathrm{P}}$ (the normalized NRS intensity) are listed in Table \ref{tab}. In definition, both $I_{\mathrm{D}}/I_{\mathrm{P}}$ and $\chi$ are proportional to the product of $n$ and $f$. In comparing the time spectra shown in Figs. \ref{ts}(c) and \ref{ts}(d), $n$ are constant, where, therefore, $I_{\mathrm{D}}/I_{\mathrm{P}}$ and $\chi$ varies only with $f$. In Table \ref{tab}, both $\chi$ and $I_{\mathrm{D}}/I_{\mathrm{P}}$ at 6 L show a large difference at 25 K and 19 K, indicating substantial reduction of $f$.

\begin{figure}[b]
\begin{center}
\includegraphics[scale=0.55, trim = 5 0 0 0]{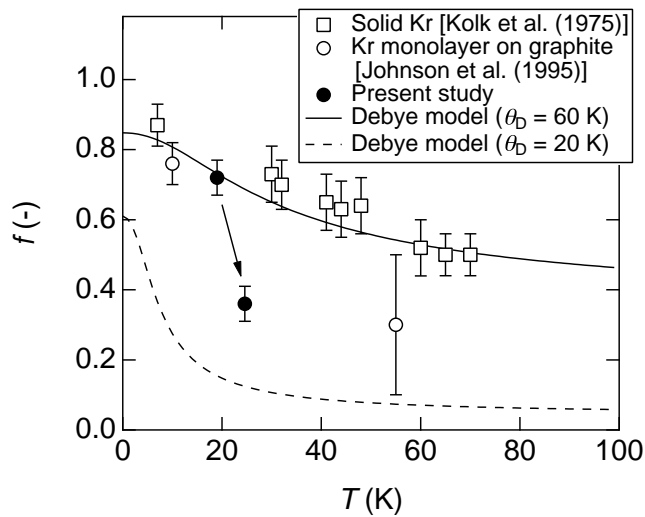} 
\caption{Temperature dependence of the recoilless fraction $f$ for solid Kr (open squares), \cite{Kolk} monolayer Kr physisorbed on graphite (open circles) \cite{Johnson} and physisorbed Kr on TiO$_{2}$(110) in the monolayer regime [present study]. \label{fvst}}
\end{center}
\end{figure}

\section{Discussion}

As shown in Table \ref{tab}, it was observed that the NRS intensity represented by $I_{\mathrm{D}}/I_{\mathrm{P}}$ and $\chi$ at 25 K sharply decreased as compared with that at 19 K. This is much larger temperature dependence at this temperature region as compared with the previous results on the solid Kr \cite{Kolk} and the monolayer Kr on the exfoliated graphite surface. \cite{Johnson} The reported values of $f$ of solid Kr at 19 K and 25 K are 0.72 and 0.70, respectively \cite{Kolk} which shows a decrease of $f$ of only about 2 \% in this temperature region. The value of $f$ of the monolayer Kr on the exfoliated graphite surface at 10 K and 55 K is $0.76\pm0.06$ and $0.3\pm0.2$, respectively. \cite{Johnson}  In both cases, $f$ shows more gradual temperature dependencies than that observed in the present study.

The possible causes of the observed temperature dependence of the NRS intensities are as follows. (i) Desorption of Kr atoms,  (ii) temperature dependence of $f$, (iii) phase transition of Kr on TiO$_{2}$(110) in the monolayer regime. We first discuss that the leading two factors should not be the major causes of the observed temperature dependence of the NRS intensity. Subsequently, we discuss the plausible mechanism of the phase transition of the Kr monolayer on the TiO$_{2}$(110) surface.

The desorption of Kr can be neglected in the present experiments both at 25 and 19 K for the following reason. In the experiment, we observed that the NRS intensity was almost constant during the irradiation of the SR beam on the sample surface at 25 and 19 K for more than 10 hours. As has been mentioned, a Kr atom in the monolayer is in general more strongly bound to the substrate surface than it does in the solid Kr. For instance, the monolayer Kr on Pt(111) is reported to desorb at $T>60$ K, \cite{Meixner} whereas the condensation of solid Kr takes place at $T<37$ K in UHV.  Below 25 K, we can safely exclude the desorption of Kr from the surface.

Next, we discuss that the observed temperature dependence of the NRS intensity is difficult to be explained by the temperature dependence of $f$ in the Debye model at any temperature range of the Debye temperature, $\theta_{\mathrm{D}}$. In the Debye model, $f$ is expressed as \cite{Goldman}
\begin{equation}
f=\exp\left[-\frac{3}{2}\frac{E_{\mathrm{R}}}{k\theta_{\mathrm{D}}}\left\{1+4\left(\frac{T}{\theta_{\mathrm{D}}}\right)^{2}\int^{\theta_{\mathrm{D}}/T}_{0}\frac{x\ dx}{e^{x-1}}\right\}\right],
\label{lmf}
\end{equation}
where the recoil energy $E_{\mathrm{R}}$ is expressed as $E_{\mathrm{R}}=(\hbar\omega_{\mathrm{e}})^{2}/2Mc^{2}$. Given that $\theta_{\mathrm{D}}=60$ K, $f$ at 19 and 25 K is calculated with Eq. (\ref{lmf}) to be 0.73 and 0.68, respectively, showing a small temperature dependence at this temperature range as shown in Fig. \ref{fvst}. Given that $\theta_{\mathrm{D}}=20$ K, $f$ at 19 and 25 K is calculated with Eq. (\ref{lmf}) to be 0.15 and 0.12, respectively, indicating that even at 20 K, the temperature dependence is still more gradual than that of the observed $I_{\mathrm{D}}/I_{\mathrm{P}}$.

The reported value of $f$ of monolayer Kr \cite{Johnson} is close to that of the solid Kr, \cite{Kolk} Therefore, we assume that the $f$ at 19 K in the monolayer Kr on TiO$_{2}$(110) is also close to the value of the solid Kr, as is depicted in Fig. \ref{fvst}. Subsequently, we deduce the relative value of $f$ at 25 K. In Fig. \ref{fvst}, one sees a noticeable deviation of the data point at 25 K from the calculated curve that exceeds the range of the error bar. Based on these two views, we conclude the conventional temperature dependence can not account for the temperature dependence of the NRS intensity observed in the present study.

\begin{figure}[t]
\begin{center}
\includegraphics[scale=1.2, trim=0 0 0 0]{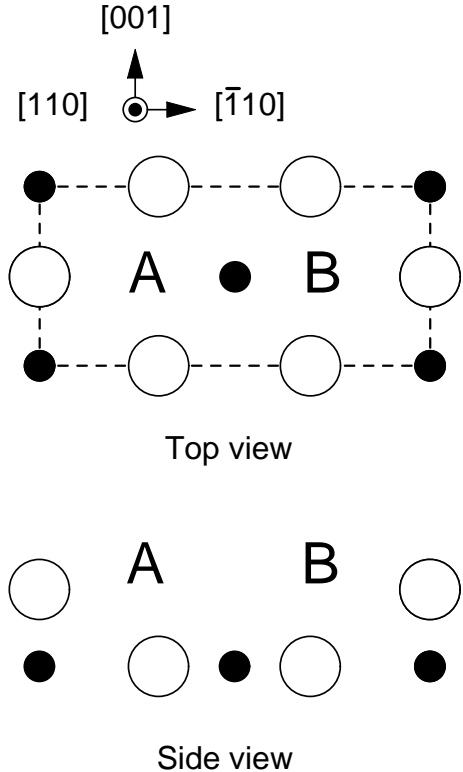} 
\caption{Unit cell of the rutile TiO$_{2}$(110) surface. Open circles and filled circles denote the O$^{2-}$ ion and Ti$^{4+}$ ion, respectively. \cite{DieboldSS} A and B denote the adsorption sites of Kr which are energetically equivalent. 
 \label{potts}}
\end{center}
\end{figure}

As a possible origin, we suggest a structural phase transition of the Kr monolayer, because a phase transition is a sudden change of the lattice structure accompanied by phonon softening, which may well sharply change $\theta_{\mathrm{D}}$, resulting in a sharp change in $f$. Here, we discuss two mechanisms of the phase transition in a physisorbed monolayer. One is the CI phase transition, \cite{Kardar} and the other is the OD phase transition. \cite{Kardar, Tabata, Matsumoto}  The CI phase transition as observed for Kr on graphite by varying the chemical potential of Kr (the gaseous pressure) and the temperature of the substrate. The CI phase transition is not likely to occur in the present system, because the gaseous pressure was fixed at UHV in the present experiment.

The OD transition may be a plausible mechanism responsible for the observed temperature dependence of $f$. In the unit cell of TiO$_{2}$(110), there are two energetically equivalent adsorption sites as depicted as A and B sites in Fig. \ref{potts} assuming that Kr prefers highly coordinated sites with oxygen anions. When the commensurate layer of physisorbed Kr is completed, one Kr atom is present in a unit cell. The situation is modeled with the spin 1/2 Ising model where the pseudo-spin variable $s_{i}$ takes 1 or -1 when Kr adsorbed on the A site or B site in each unit cell, respectively, where the Hamiltonian is described as
\begin{equation}
\mathcal{H}=-\sum_{ij}J_{ij} s_{i}s_{j}.
\label{ising}
\end{equation}
where the interaction energy $J_{ij}$ is determined by the competition between the van der Waals interaction and repulsion between the induced dipoles. In the $[001]$ direction, $J_{ij}$ for the nearest neighboring Kr atoms is expected to be negative, for the inter Kr separation is small where the Pauli repulsion term should dominate, resulting in the formation of the antiferromagnetically ordered chain in this direction. On the other hand, $J_{ij}$ in the $[\bar{1}10]$ direction may be positive or negative, for the attractive dispersion force and the repulsive dipole interaction compete in this range of the inter Kr separation.

With the above anisotropic model, we expect an OD phase transition by analogy to the OD transition observed on the Si(001) surface. \cite{Tabata, Matsumoto} Here, we assume that the critical temperature $(T_{\mathrm{C}})$ is located between 19 and 25 K to account for the observed temperature dependence of $f$. At $T<T_{\mathrm{C}}$, the ferromagnetically or antiferromagnetically ordered structure in the $[\bar{1}10]$ direction is formed, whereas the structure becomes disordered at $T>T_{\mathrm{C}}$. In the disordered phase, the Kr atom flips its position between the A and B sites rather freely in a unit cell, which will result in the softening of the Kr lattice in $[\bar{1}10]$ direction in which the SR was incident on the surface. We therefore propose the temperature dependence of the NRS intensity observed in the present study may be a consequence of this structural change which leads to the decrease of $\theta_{D}$ in the $[\bar{1}10]$ direction. Further structural and spectroscopic studies will help to clarify the above anticipation.

\section{Conclusion}
The NRS from $^{83}$Kr was obtained for Kr layers with various thicknesses on TiO$_{2}$(110). 
The NRS signal corresponding to monolayer was estimated to be 0.23 cps. 
The time spectra were analyzed by taking account of the dynamical effect, and the recoilless fraction was evaluated. 
For the Kr monolayer, the NRS intensity and the recoilless fraction were found to sharply decrease upon temperature rise from 19 to 25 K. It is discussed that the OD phase transition of the Kr monolayer occurs between 19 and 25 K inducing softening of the Kr lattice in the $[\bar{1}10]$ direction.

\section{Aknowledgement}
This work was supported by Grant-in Aid for Scientific Research (Grant Nos. 11J09249, 24246013 and 24108503) of Japan Society for the Promotion of Science (JSPS). 
This work was performed under the approval of the proposals 2011B1188, 2011B1695, 2012A1560, 2012B1699 and 2013A1717 by Japan Synchrotron Radiation Research Institute (JASRI). 
AI acknowledges the support from the program ''Budding researchers support proposal'' by SPring-8.

% Create the reference section using BibTeX:
% \bibliography{mybibfile}

\begin{thebibliography}{41}%
\makeatletter
\providecommand \@ifxundefined [1]{%
 \@ifx{#1\undefined}
}%
\providecommand \@ifnum [1]{%
 \ifnum #1\expandafter \@firstoftwo
 \else \expandafter \@secondoftwo
 \fi
}%
\providecommand \@ifx [1]{%
 \ifx #1\expandafter \@firstoftwo
 \else \expandafter \@secondoftwo
 \fi
}%
\providecommand \natexlab [1]{#1}%
\providecommand \enquote  [1]{``#1''}%
\providecommand \bibnamefont  [1]{#1}%
\providecommand \bibfnamefont [1]{#1}%
\providecommand \citenamefont [1]{#1}%
\providecommand \href@noop [0]{\@secondoftwo}%
\providecommand \href [0]{\begingroup \@sanitize@url \@href}%
\providecommand \@href[1]{\@@startlink{#1}\@@href}%
\providecommand \@@href[1]{\endgroup#1\@@endlink}%
\providecommand \@sanitize@url [0]{\catcode `\\12\catcode `\$12\catcode
  `\&12\catcode `\#12\catcode `\^12\catcode `\_12\catcode `\%12\relax}%
\providecommand \@@startlink[1]{}%
\providecommand \@@endlink[0]{}%
\providecommand \url  [0]{\begingroup\@sanitize@url \@url }%
\providecommand \@url [1]{\endgroup\@href {#1}{\urlprefix }}%
\providecommand \urlprefix  [0]{URL }%
\providecommand \Eprint [0]{\href }%
\providecommand \doibase [0]{http://dx.doi.org/}%
\providecommand \selectlanguage [0]{\@gobble}%
\providecommand \bibinfo  [0]{\@secondoftwo}%
\providecommand \bibfield  [0]{\@secondoftwo}%
\providecommand \translation [1]{[#1]}%
\providecommand \BibitemOpen [0]{}%
\providecommand \bibitemStop [0]{}%
\providecommand \bibitemNoStop [0]{.\EOS\space}%
\providecommand \EOS [0]{\spacefactor3000\relax}%
\providecommand \BibitemShut  [1]{\csname bibitem#1\endcsname}%
\let\auto@bib@innerbib\@empty
%</preamble>
\bibitem [{\citenamefont {R\"ohlsberger}(2004)}]{Rohlsberger}%
  \BibitemOpen
  \bibfield  {author} {\bibinfo {author} {\bibfnamefont {R.}~\bibnamefont
  {R\"ohlsberger}},\ }\href@noop {} {\emph {\bibinfo {title} {\textit{Nuclear
  Condensed Matter Physics with Synchrotron Radiation}}}}\ (\bibinfo
  {publisher} {Springer-Verlag},\ \bibinfo {address} {Berlin},\ \bibinfo {year}
  {2004})\BibitemShut {NoStop}%
\bibitem [{\citenamefont {Baron}\ \emph {et~al.}(1995)\citenamefont {Baron},
  \citenamefont {Chumakov}, \citenamefont {Ruby}, \citenamefont {Arthur},
  \citenamefont {Brown}, \citenamefont {Smirnov},\ and\ \citenamefont {van
  B\"urck}}]{Baron}%
  \BibitemOpen
  \bibfield  {author} {\bibinfo {author} {\bibfnamefont {A.~Q.~R.}\
  \bibnamefont {Baron}}, \bibinfo {author} {\bibfnamefont {A.~I.}\ \bibnamefont
  {Chumakov}}, \bibinfo {author} {\bibfnamefont {S.~L.}\ \bibnamefont {Ruby}},
  \bibinfo {author} {\bibfnamefont {J.}~\bibnamefont {Arthur}}, \bibinfo
  {author} {\bibfnamefont {G.~S.}\ \bibnamefont {Brown}}, \bibinfo {author}
  {\bibfnamefont {G.~V.}\ \bibnamefont {Smirnov}}, \ and\ \bibinfo {author}
  {\bibfnamefont {U.}~\bibnamefont {van B\"urck}},\ }\href@noop {} {\bibfield
  {journal} {\bibinfo  {journal} {Phys. Rev. B}\ }\textbf {\bibinfo {volume}
  {51}},\ \bibinfo {pages} {16384} (\bibinfo {year} {1995})}\BibitemShut
  {NoStop}%
\bibitem [{\citenamefont {Klobes}\ \emph {et~al.}(2013)\citenamefont {Klobes},
  \citenamefont {Desmedt}, \citenamefont {Sergueev}, \citenamefont {Schmalzl},\
  and\ \citenamefont {Hermann}}]{Klobes}%
  \BibitemOpen
  \bibfield  {author} {\bibinfo {author} {\bibfnamefont {B.}~\bibnamefont
  {Klobes}}, \bibinfo {author} {\bibfnamefont {A.}~\bibnamefont {Desmedt}},
  \bibinfo {author} {\bibfnamefont {I.}~\bibnamefont {Sergueev}}, \bibinfo
  {author} {\bibfnamefont {K.}~\bibnamefont {Schmalzl}}, \ and\ \bibinfo
  {author} {\bibfnamefont {R.~P.}\ \bibnamefont {Hermann}},\ }\href@noop {}
  {\bibfield  {journal} {\bibinfo  {journal} {Europhys. Lett.}\ }\textbf
  {\bibinfo {volume} {103}},\ \bibinfo {pages} {36001} (\bibinfo {year}
  {2013})}\BibitemShut {NoStop}%
\bibitem [{\citenamefont {Kishimoto}\ \emph {et~al.}(2000)\citenamefont
  {Kishimoto}, \citenamefont {Yoda}, \citenamefont {Seto}, \citenamefont
  {Kobayashi}, \citenamefont {Kitao}, \citenamefont {Haruki}, \citenamefont
  {Kawauchi}, \citenamefont {Fukutani},\ and\ \citenamefont
  {Okano}}]{Kishimoto}%
  \BibitemOpen
  \bibfield  {author} {\bibinfo {author} {\bibfnamefont {S.}~\bibnamefont
  {Kishimoto}}, \bibinfo {author} {\bibfnamefont {Y.}~\bibnamefont {Yoda}},
  \bibinfo {author} {\bibfnamefont {M.}~\bibnamefont {Seto}}, \bibinfo {author}
  {\bibfnamefont {Y.}~\bibnamefont {Kobayashi}}, \bibinfo {author}
  {\bibfnamefont {S.}~\bibnamefont {Kitao}}, \bibinfo {author} {\bibfnamefont
  {R.}~\bibnamefont {Haruki}}, \bibinfo {author} {\bibfnamefont
  {T.}~\bibnamefont {Kawauchi}}, \bibinfo {author} {\bibfnamefont
  {K.}~\bibnamefont {Fukutani}}, \ and\ \bibinfo {author} {\bibfnamefont
  {T.}~\bibnamefont {Okano}},\ }\href@noop {} {\bibfield  {journal} {\bibinfo
  {journal} {Phys. Rev. Lett.}\ }\textbf {\bibinfo {volume} {85}},\ \bibinfo
  {pages} {1831} (\bibinfo {year} {2000})}\BibitemShut {NoStop}%
\bibitem [{\citenamefont {Matsuoka}\ \emph {et~al.}(2011)\citenamefont
  {Matsuoka}, \citenamefont {Fujihisa}, \citenamefont {Hirao}, \citenamefont
  {Ohishi}, \citenamefont {Mitsui}, \citenamefont {Masuda}, \citenamefont
  {Seto}, \citenamefont {Yoda}, \citenamefont {Shimizu}, \citenamefont
  {Machida},\ and\ \citenamefont {Aoki}}]{Matsuoka}%
  \BibitemOpen
  \bibfield  {author} {\bibinfo {author} {\bibfnamefont {T.}~\bibnamefont
  {Matsuoka}}, \bibinfo {author} {\bibfnamefont {H.}~\bibnamefont {Fujihisa}},
  \bibinfo {author} {\bibfnamefont {N.}~\bibnamefont {Hirao}}, \bibinfo
  {author} {\bibfnamefont {Y.}~\bibnamefont {Ohishi}}, \bibinfo {author}
  {\bibfnamefont {T.}~\bibnamefont {Mitsui}}, \bibinfo {author} {\bibfnamefont
  {R.}~\bibnamefont {Masuda}}, \bibinfo {author} {\bibfnamefont
  {M.}~\bibnamefont {Seto}}, \bibinfo {author} {\bibfnamefont {Y.}~\bibnamefont
  {Yoda}}, \bibinfo {author} {\bibfnamefont {K.}~\bibnamefont {Shimizu}},
  \bibinfo {author} {\bibfnamefont {A.}~\bibnamefont {Machida}}, \ and\
  \bibinfo {author} {\bibfnamefont {K.}~\bibnamefont {Aoki}},\ }\href@noop {}
  {\bibfield  {journal} {\bibinfo  {journal} {Phys. Rev. Lett.}\ }\textbf
  {\bibinfo {volume} {107}},\ \bibinfo {pages} {025501} (\bibinfo {year}
  {2011})}\BibitemShut {NoStop}%
\bibitem [{\citenamefont {Kawauchi}\ \emph {et~al.}(2011)\citenamefont
  {Kawauchi}, \citenamefont {Fukutani}, \citenamefont {Matsumoto},
  \citenamefont {Oda}, \citenamefont {Okano}, \citenamefont {Zhang},
  \citenamefont {Kishimoto},\ and\ \citenamefont {Yoda}}]{Kawauchi1}%
  \BibitemOpen
  \bibfield  {author} {\bibinfo {author} {\bibfnamefont {T.}~\bibnamefont
  {Kawauchi}}, \bibinfo {author} {\bibfnamefont {K.}~\bibnamefont {Fukutani}},
  \bibinfo {author} {\bibfnamefont {M.}~\bibnamefont {Matsumoto}}, \bibinfo
  {author} {\bibfnamefont {K.}~\bibnamefont {Oda}}, \bibinfo {author}
  {\bibfnamefont {T.}~\bibnamefont {Okano}}, \bibinfo {author} {\bibfnamefont
  {X.~W.}\ \bibnamefont {Zhang}}, \bibinfo {author} {\bibfnamefont
  {S.}~\bibnamefont {Kishimoto}}, \ and\ \bibinfo {author} {\bibfnamefont
  {Y.}~\bibnamefont {Yoda}},\ }\href@noop {} {\bibfield  {journal} {\bibinfo
  {journal} {Phys. Rev. B}\ }\textbf {\bibinfo {volume} {84}},\ \bibinfo
  {pages} {020415} (\bibinfo {year} {2011})}\BibitemShut {NoStop}%
\bibitem [{\citenamefont {\'{S}l\c{e}zak}\ \emph
  {et~al.}(2013{\natexlab{a}})\citenamefont {\'{S}l\c{e}zak}, \citenamefont
  {\'{S}l\c{e}zak}, \citenamefont {Freindl}, \citenamefont {Kara\'{s}},
  \citenamefont {Spiridis}, \citenamefont {Zaj\c{a}c}, \citenamefont
  {Chumakov}, \citenamefont {Stankov}, \citenamefont {R\"uffer},\ and\
  \citenamefont {Korecki}}]{Slezak1}%
  \BibitemOpen
  \bibfield  {author} {\bibinfo {author} {\bibfnamefont {M.}~\bibnamefont
  {\'{S}l\c{e}zak}}, \bibinfo {author} {\bibfnamefont {T.}~\bibnamefont
  {\'{S}l\c{e}zak}}, \bibinfo {author} {\bibfnamefont {K.}~\bibnamefont
  {Freindl}}, \bibinfo {author} {\bibfnamefont {W.}~\bibnamefont {Kara\'{s}}},
  \bibinfo {author} {\bibfnamefont {N.}~\bibnamefont {Spiridis}}, \bibinfo
  {author} {\bibfnamefont {M.}~\bibnamefont {Zaj\c{a}c}}, \bibinfo {author}
  {\bibfnamefont {A.~I.}\ \bibnamefont {Chumakov}}, \bibinfo {author}
  {\bibfnamefont {S.}~\bibnamefont {Stankov}}, \bibinfo {author} {\bibfnamefont
  {R.}~\bibnamefont {R\"uffer}}, \ and\ \bibinfo {author} {\bibfnamefont
  {J.}~\bibnamefont {Korecki}},\ }\href@noop {} {\bibfield  {journal} {\bibinfo
   {journal} {Phys. Rev. B}\ }\textbf {\bibinfo {volume} {87}},\ \bibinfo
  {pages} {134411} (\bibinfo {year} {2013}{\natexlab{a}})}\BibitemShut
  {NoStop}%
\bibitem [{\citenamefont {\'{S}l\c{e}zak}\ \emph
  {et~al.}(2013{\natexlab{b}})\citenamefont {\'{S}l\c{e}zak}, \citenamefont
  {Zaj\c{a}c}, \citenamefont {\'{S}l\c{e}zak}, \citenamefont {Matlak},
  \citenamefont {Kozio\l-Rachwa\l}, \citenamefont {Wilgocka-\'{S}l\c{e}zak},
  \citenamefont {Chumakov}, \citenamefont {R\"uffer},\ and\ \citenamefont
  {Korecki}}]{Slezak2}%
  \BibitemOpen
  \bibfield  {author} {\bibinfo {author} {\bibfnamefont {T.}~\bibnamefont
  {\'{S}l\c{e}zak}}, \bibinfo {author} {\bibfnamefont {M.}~\bibnamefont
  {Zaj\c{a}c}}, \bibinfo {author} {\bibfnamefont {M.}~\bibnamefont
  {\'{S}l\c{e}zak}}, \bibinfo {author} {\bibfnamefont {K.}~\bibnamefont
  {Matlak}}, \bibinfo {author} {\bibfnamefont {A.}~\bibnamefont
  {Kozio\l-Rachwa\l}}, \bibinfo {author} {\bibfnamefont {D.}~\bibnamefont
  {Wilgocka-\'{S}l\c{e}zak}}, \bibinfo {author} {\bibfnamefont {A.~I.}\
  \bibnamefont {Chumakov}}, \bibinfo {author} {\bibfnamefont {R.}~\bibnamefont
  {R\"uffer}}, \ and\ \bibinfo {author} {\bibfnamefont {J.}~\bibnamefont
  {Korecki}},\ }\href@noop {} {\bibfield  {journal} {\bibinfo  {journal} {Phys.
  Rev. B}\ }\textbf {\bibinfo {volume} {87}},\ \bibinfo {pages} {094423}
  (\bibinfo {year} {2013}{\natexlab{b}})}\BibitemShut {NoStop}%
\bibitem [{\citenamefont {R\"ohlsberger}\ \emph {et~al.}(2001)\citenamefont
  {R\"ohlsberger}, \citenamefont {Bansmann}, \citenamefont {Senz},
  \citenamefont {Jonas}, \citenamefont {Bettac}, \citenamefont {Leupold},
  \citenamefont {R\"uffer}, \citenamefont {Burkel},\ and\ \citenamefont
  {Meiwes-Broer}}]{RohlsbergerPRL}%
  \BibitemOpen
  \bibfield  {author} {\bibinfo {author} {\bibfnamefont {R.}~\bibnamefont
  {R\"ohlsberger}}, \bibinfo {author} {\bibfnamefont {J.}~\bibnamefont
  {Bansmann}}, \bibinfo {author} {\bibfnamefont {V.}~\bibnamefont {Senz}},
  \bibinfo {author} {\bibfnamefont {K.~L.}\ \bibnamefont {Jonas}}, \bibinfo
  {author} {\bibfnamefont {A.}~\bibnamefont {Bettac}}, \bibinfo {author}
  {\bibfnamefont {O.}~\bibnamefont {Leupold}}, \bibinfo {author} {\bibfnamefont
  {R.}~\bibnamefont {R\"uffer}}, \bibinfo {author} {\bibfnamefont
  {E.}~\bibnamefont {Burkel}}, \ and\ \bibinfo {author} {\bibfnamefont {K.~H.}\
  \bibnamefont {Meiwes-Broer}},\ }\href@noop {} {\bibfield  {journal} {\bibinfo
   {journal} {Phys. Rev. Lett.}\ }\textbf {\bibinfo {volume} {86}},\ \bibinfo
  {pages} {5597} (\bibinfo {year} {2001})}\BibitemShut {NoStop}%
\bibitem [{\citenamefont {Sladecek}\ \emph {et~al.}(2004)\citenamefont
  {Sladecek}, \citenamefont {Sepiol}, \citenamefont {Korecki}, \citenamefont
  {Slezak}, \citenamefont {R\"uffer}, \citenamefont {Kmiec},\ and\
  \citenamefont {Vogl}}]{Sladecek}%
  \BibitemOpen
  \bibfield  {author} {\bibinfo {author} {\bibfnamefont {M.}~\bibnamefont
  {Sladecek}}, \bibinfo {author} {\bibfnamefont {B.}~\bibnamefont {Sepiol}},
  \bibinfo {author} {\bibfnamefont {J.}~\bibnamefont {Korecki}}, \bibinfo
  {author} {\bibfnamefont {T.}~\bibnamefont {Slezak}}, \bibinfo {author}
  {\bibfnamefont {R.}~\bibnamefont {R\"uffer}}, \bibinfo {author}
  {\bibfnamefont {D.}~\bibnamefont {Kmiec}}, \ and\ \bibinfo {author}
  {\bibfnamefont {G.}~\bibnamefont {Vogl}},\ }\href@noop {} {\bibfield
  {journal} {\bibinfo  {journal} {Surf. Sci.}\ }\textbf {\bibinfo {volume}
  {566-568}},\ \bibinfo {pages} {372} (\bibinfo {year} {2004})}\BibitemShut
  {NoStop}%
\bibitem [{\citenamefont {Vogl}\ \emph {et~al.}(2007)\citenamefont {Vogl},
  \citenamefont {Sladecek},\ and\ \citenamefont {Dattagupta}}]{Vogl2007}%
  \BibitemOpen
  \bibfield  {author} {\bibinfo {author} {\bibfnamefont {G.}~\bibnamefont
  {Vogl}}, \bibinfo {author} {\bibfnamefont {M.}~\bibnamefont {Sladecek}}, \
  and\ \bibinfo {author} {\bibfnamefont {S.}~\bibnamefont {Dattagupta}},\
  }\href@noop {} {\bibfield  {journal} {\bibinfo  {journal} {Phys. Rev. Lett.}\
  }\textbf {\bibinfo {volume} {99}},\ \bibinfo {pages} {155902} (\bibinfo
  {year} {2007})}\BibitemShut {NoStop}%
\bibitem [{\citenamefont {Partykajankowska}\ \emph {et~al.}(2008)\citenamefont
  {Partykajankowska}, \citenamefont {Sepiol}, \citenamefont {Sladecek},
  \citenamefont {Kmiec}, \citenamefont {Korecki}, \citenamefont {Slezak},
  \citenamefont {Zajac}, \citenamefont {Stankov}, \citenamefont {R\"uffer},\
  and\ \citenamefont {Vogl}}]{Partykajankowska}%
  \BibitemOpen
  \bibfield  {author} {\bibinfo {author} {\bibfnamefont {E.}~\bibnamefont
  {Partykajankowska}}, \bibinfo {author} {\bibfnamefont {B.}~\bibnamefont
  {Sepiol}}, \bibinfo {author} {\bibfnamefont {M.}~\bibnamefont {Sladecek}},
  \bibinfo {author} {\bibfnamefont {D.}~\bibnamefont {Kmiec}}, \bibinfo
  {author} {\bibfnamefont {J.}~\bibnamefont {Korecki}}, \bibinfo {author}
  {\bibfnamefont {T.}~\bibnamefont {Slezak}}, \bibinfo {author} {\bibfnamefont
  {M.}~\bibnamefont {Zajac}}, \bibinfo {author} {\bibfnamefont
  {S.}~\bibnamefont {Stankov}}, \bibinfo {author} {\bibfnamefont
  {R.}~\bibnamefont {R\"uffer}}, \ and\ \bibinfo {author} {\bibfnamefont
  {G.}~\bibnamefont {Vogl}},\ }\href@noop {} {\bibfield  {journal} {\bibinfo
  {journal} {Surf. Sci.}\ }\textbf {\bibinfo {volume} {602}},\ \bibinfo {pages}
  {1453} (\bibinfo {year} {2008})}\BibitemShut {NoStop}%
\bibitem [{\citenamefont {Vogl}\ \emph {et~al.}(2009)\citenamefont {Vogl},
  \citenamefont {Partyka-Jankowska}, \citenamefont {Zaj\c{a}c},\ and\
  \citenamefont {Chumakov}}]{Vogl2009}%
  \BibitemOpen
  \bibfield  {author} {\bibinfo {author} {\bibfnamefont {G.}~\bibnamefont
  {Vogl}}, \bibinfo {author} {\bibfnamefont {E.}~\bibnamefont
  {Partyka-Jankowska}}, \bibinfo {author} {\bibfnamefont {M.}~\bibnamefont
  {Zaj\c{a}c}}, \ and\ \bibinfo {author} {\bibfnamefont {A.~I.}\ \bibnamefont
  {Chumakov}},\ }\href@noop {} {\bibfield  {journal} {\bibinfo  {journal}
  {Phys. Rev. B}\ }\textbf {\bibinfo {volume} {80}},\ \bibinfo {pages} {115406}
  (\bibinfo {year} {2009})}\BibitemShut {NoStop}%
\bibitem [{\citenamefont {Kardar}\ and\ \citenamefont {Berker}(1982)}]{Kardar}%
  \BibitemOpen
  \bibfield  {author} {\bibinfo {author} {\bibfnamefont {M.}~\bibnamefont
  {Kardar}}\ and\ \bibinfo {author} {\bibfnamefont {A.~N.}\ \bibnamefont
  {Berker}},\ }\href@noop {} {\bibfield  {journal} {\bibinfo  {journal} {Phys.
  Rev. Lett.}\ }\textbf {\bibinfo {volume} {48}},\ \bibinfo {pages} {1552}
  (\bibinfo {year} {1982})}\BibitemShut {NoStop}%
\bibitem [{\citenamefont {Wandelt}(1984)}]{Wandelt}%
  \BibitemOpen
  \bibfield  {author} {\bibinfo {author} {\bibfnamefont {K.}~\bibnamefont
  {Wandelt}},\ }\href@noop {} {\bibfield  {journal} {\bibinfo  {journal} {J.
  Vac. Sci. Technol. A}\ }\textbf {\bibinfo {volume} {2}},\ \bibinfo {pages}
  {802} (\bibinfo {year} {1984})}\BibitemShut {NoStop}%
\bibitem [{\citenamefont {Getzlaff}\ \emph {et~al.}(1993)\citenamefont
  {Getzlaff}, \citenamefont {Bansmann},\ and\ \citenamefont
  {Sch\"onhense}}]{Getzlaff}%
  \BibitemOpen
  \bibfield  {author} {\bibinfo {author} {\bibfnamefont {M.}~\bibnamefont
  {Getzlaff}}, \bibinfo {author} {\bibfnamefont {J.}~\bibnamefont {Bansmann}},
  \ and\ \bibinfo {author} {\bibfnamefont {G.}~\bibnamefont {Sch\"onhense}},\
  }\href@noop {} {\bibfield  {journal} {\bibinfo  {journal} {Phys. Rev. Lett.}\
  }\textbf {\bibinfo {volume} {71}},\ \bibinfo {pages} {793} (\bibinfo {year}
  {1993})}\BibitemShut {NoStop}%
\bibitem [{\citenamefont {Guo}\ and\ \citenamefont {Zaera}(2006)}]{Guo}%
  \BibitemOpen
  \bibfield  {author} {\bibinfo {author} {\bibfnamefont {H.}~\bibnamefont
  {Guo}}\ and\ \bibinfo {author} {\bibfnamefont {F.}~\bibnamefont {Zaera}},\
  }\href@noop {} {\bibfield  {journal} {\bibinfo  {journal} {Nature Mat.}\
  }\textbf {\bibinfo {volume} {5}},\ \bibinfo {pages} {489} (\bibinfo {year}
  {2006})}\BibitemShut {NoStop}%
\bibitem [{\citenamefont {Bellini}\ \emph {et~al.}(2004)\citenamefont
  {Bellini}, \citenamefont {Cottenier}, \citenamefont {\c{C}akmak},
  \citenamefont {Manghi},\ and\ \citenamefont {Rots}}]{Bellini}%
  \BibitemOpen
  \bibfield  {author} {\bibinfo {author} {\bibfnamefont {V.}~\bibnamefont
  {Bellini}}, \bibinfo {author} {\bibfnamefont {S.}~\bibnamefont {Cottenier}},
  \bibinfo {author} {\bibfnamefont {M.}~\bibnamefont {\c{C}akmak}}, \bibinfo
  {author} {\bibfnamefont {F.}~\bibnamefont {Manghi}}, \ and\ \bibinfo {author}
  {\bibfnamefont {M.}~\bibnamefont {Rots}},\ }\href@noop {} {\bibfield
  {journal} {\bibinfo  {journal} {Phys. Rev. B}\ }\textbf {\bibinfo {volume}
  {70}},\ \bibinfo {pages} {155419} (\bibinfo {year} {2004})}\BibitemShut
  {NoStop}%
\bibitem [{\citenamefont {Cottenier}\ \emph {et~al.}(2004)\citenamefont
  {Cottenier}, \citenamefont {Bellini}, \citenamefont {\c{C}akmak},
  \citenamefont {Manghi},\ and\ \citenamefont {Rots}}]{Cottenier}%
  \BibitemOpen
  \bibfield  {author} {\bibinfo {author} {\bibfnamefont {S.}~\bibnamefont
  {Cottenier}}, \bibinfo {author} {\bibfnamefont {V.}~\bibnamefont {Bellini}},
  \bibinfo {author} {\bibfnamefont {M.}~\bibnamefont {\c{C}akmak}}, \bibinfo
  {author} {\bibfnamefont {F.}~\bibnamefont {Manghi}}, \ and\ \bibinfo {author}
  {\bibfnamefont {M.}~\bibnamefont {Rots}},\ }\href@noop {} {\bibfield
  {journal} {\bibinfo  {journal} {Phys. Rev. B}\ }\textbf {\bibinfo {volume}
  {70}},\ \bibinfo {pages} {155418} (\bibinfo {year} {2004})}\BibitemShut
  {NoStop}%
\bibitem [{\citenamefont {Ikeda}\ \emph {et~al.}(2011)\citenamefont {Ikeda},
  \citenamefont {Matsumoto}, \citenamefont {Ogura}, \citenamefont {Fukutani},\
  and\ \citenamefont {Okano}}]{Ikeda1}%
  \BibitemOpen
  \bibfield  {author} {\bibinfo {author} {\bibfnamefont {A.}~\bibnamefont
  {Ikeda}}, \bibinfo {author} {\bibfnamefont {M.}~\bibnamefont {Matsumoto}},
  \bibinfo {author} {\bibfnamefont {S.}~\bibnamefont {Ogura}}, \bibinfo
  {author} {\bibfnamefont {K.}~\bibnamefont {Fukutani}}, \ and\ \bibinfo
  {author} {\bibfnamefont {T.}~\bibnamefont {Okano}},\ }\href@noop {}
  {\bibfield  {journal} {\bibinfo  {journal} {Phys. Rev. B}\ }\textbf {\bibinfo
  {volume} {84}},\ \bibinfo {pages} {155412} (\bibinfo {year}
  {2011})}\BibitemShut {NoStop}%
\bibitem [{\citenamefont {Bagus}\ \emph {et~al.}(2002)\citenamefont {Bagus},
  \citenamefont {Staemmler},\ and\ \citenamefont {Woll}}]{Bagus}%
  \BibitemOpen
  \bibfield  {author} {\bibinfo {author} {\bibfnamefont {P.~S.}\ \bibnamefont
  {Bagus}}, \bibinfo {author} {\bibfnamefont {V.}~\bibnamefont {Staemmler}}, \
  and\ \bibinfo {author} {\bibfnamefont {C.}~\bibnamefont {Woll}},\ }\href@noop
  {} {\bibfield  {journal} {\bibinfo  {journal} {Phys. Rev. Lett.}\ }\textbf
  {\bibinfo {volume} {89}},\ \bibinfo {pages} {096104} (\bibinfo {year}
  {2002})}\BibitemShut {NoStop}%
\bibitem [{\citenamefont {Da~Silva}\ and\ \citenamefont
  {Stampfl}(2008)}]{Silva1}%
  \BibitemOpen
  \bibfield  {author} {\bibinfo {author} {\bibfnamefont {J.~L.~F.}\
  \bibnamefont {Da~Silva}}\ and\ \bibinfo {author} {\bibfnamefont
  {C.}~\bibnamefont {Stampfl}},\ }\href@noop {} {\bibfield  {journal} {\bibinfo
   {journal} {Phys. Rev. B}\ }\textbf {\bibinfo {volume} {77}},\ \bibinfo
  {pages} {045401} (\bibinfo {year} {2008})}\BibitemShut {NoStop}%
\bibitem [{\citenamefont {Chen}\ \emph {et~al.}(2011)\citenamefont {Chen},
  \citenamefont {Al-Saidi},\ and\ \citenamefont {Johnson}}]{Chen}%
  \BibitemOpen
  \bibfield  {author} {\bibinfo {author} {\bibfnamefont {D.-L.}\ \bibnamefont
  {Chen}}, \bibinfo {author} {\bibfnamefont {W.~A.}\ \bibnamefont {Al-Saidi}},
  \ and\ \bibinfo {author} {\bibfnamefont {J.~K.}\ \bibnamefont {Johnson}},\
  }\href@noop {} {\bibfield  {journal} {\bibinfo  {journal} {Phys. Rev. B}\
  }\textbf {\bibinfo {volume} {84}},\ \bibinfo {pages} {241405} (\bibinfo
  {year} {2011})}\BibitemShut {NoStop}%
\bibitem [{\citenamefont {Johnson}\ \emph {et~al.}(1995)\citenamefont
  {Johnson}, \citenamefont {Siddons}, \citenamefont {Larese},\ and\
  \citenamefont {Hastings}}]{Johnson}%
  \BibitemOpen
  \bibfield  {author} {\bibinfo {author} {\bibfnamefont {D.~E.}\ \bibnamefont
  {Johnson}}, \bibinfo {author} {\bibfnamefont {D.~P.}\ \bibnamefont
  {Siddons}}, \bibinfo {author} {\bibfnamefont {J.~Z.}\ \bibnamefont {Larese}},
  \ and\ \bibinfo {author} {\bibfnamefont {J.~B.}\ \bibnamefont {Hastings}},\
  }\href@noop {} {\bibfield  {journal} {\bibinfo  {journal} {Phys. Rev. B}\
  }\textbf {\bibinfo {volume} {51}},\ \bibinfo {pages} {7909} (\bibinfo {year}
  {1995})}\BibitemShut {NoStop}%
\bibitem [{\citenamefont {Tabata}\ \emph {et~al.}(1987)\citenamefont {Tabata},
  \citenamefont {Aruga},\ and\ \citenamefont {Murata}}]{Tabata}%
  \BibitemOpen
  \bibfield  {author} {\bibinfo {author} {\bibfnamefont {T.}~\bibnamefont
  {Tabata}}, \bibinfo {author} {\bibfnamefont {T.}~\bibnamefont {Aruga}}, \
  and\ \bibinfo {author} {\bibfnamefont {Y.}~\bibnamefont {Murata}},\
  }\href@noop {} {\bibfield  {journal} {\bibinfo  {journal} {Surf. Sci.}\
  }\textbf {\bibinfo {volume} {179}},\ \bibinfo {pages} {L63} (\bibinfo {year}
  {1987})}\BibitemShut {NoStop}%
\bibitem [{\citenamefont {Matsumoto}\ \emph {et~al.}(2003)\citenamefont
  {Matsumoto}, \citenamefont {Fukutani},\ and\ \citenamefont
  {Okano}}]{Matsumoto}%
  \BibitemOpen
  \bibfield  {author} {\bibinfo {author} {\bibfnamefont {M.}~\bibnamefont
  {Matsumoto}}, \bibinfo {author} {\bibfnamefont {K.}~\bibnamefont {Fukutani}},
  \ and\ \bibinfo {author} {\bibfnamefont {T.}~\bibnamefont {Okano}},\
  }\href@noop {} {\bibfield  {journal} {\bibinfo  {journal} {Phys. Rev. Lett.}\
  }\textbf {\bibinfo {volume} {90}},\ \bibinfo {pages} {106103} (\bibinfo
  {year} {2003})}\BibitemShut {NoStop}%
\bibitem [{\citenamefont {Kolk}(1975)}]{Kolk}%
  \BibitemOpen
  \bibfield  {author} {\bibinfo {author} {\bibfnamefont {B.}~\bibnamefont
  {Kolk}},\ }\href@noop {} {\bibfield  {journal} {\bibinfo  {journal} {Phys.
  Rev. B}\ }\textbf {\bibinfo {volume} {12}},\ \bibinfo {pages} {4695}
  (\bibinfo {year} {1975})}\BibitemShut {NoStop}%
\bibitem [{\citenamefont {Kawauchi}\ \emph {et~al.}(2009)\citenamefont
  {Kawauchi}, \citenamefont {Matsumoto}, \citenamefont {Fukutani},
  \citenamefont {Okano}, \citenamefont {Suetsugu}, \citenamefont {Zhang},\ and\
  \citenamefont {Yoda}}]{Kawauchi2}%
  \BibitemOpen
  \bibfield  {author} {\bibinfo {author} {\bibfnamefont {T.}~\bibnamefont
  {Kawauchi}}, \bibinfo {author} {\bibfnamefont {M.}~\bibnamefont {Matsumoto}},
  \bibinfo {author} {\bibfnamefont {K.}~\bibnamefont {Fukutani}}, \bibinfo
  {author} {\bibfnamefont {T.}~\bibnamefont {Okano}}, \bibinfo {author}
  {\bibfnamefont {Y.}~\bibnamefont {Suetsugu}}, \bibinfo {author}
  {\bibfnamefont {X.~W.}\ \bibnamefont {Zhang}}, \ and\ \bibinfo {author}
  {\bibfnamefont {Y.}~\bibnamefont {Yoda}},\ }\href@noop {} {\bibfield
  {journal} {\bibinfo  {journal} {Vacuum}\ }\textbf {\bibinfo {volume} {83}},\
  \bibinfo {pages} {873} (\bibinfo {year} {2009})}\BibitemShut {NoStop}%
\bibitem [{\citenamefont {Fukada}\ \emph {et~al.}()\citenamefont {Fukada},
  \citenamefont {Matsumoto}, \citenamefont {Takeyasu}, \citenamefont {Ogura},\
  and\ \citenamefont {Fukutani}}]{Fukada}%
  \BibitemOpen
  \bibfield  {author} {\bibinfo {author} {\bibfnamefont {K.}~\bibnamefont
  {Fukada}}, \bibinfo {author} {\bibfnamefont {M.}~\bibnamefont {Matsumoto}},
  \bibinfo {author} {\bibfnamefont {K.}~\bibnamefont {Takeyasu}}, \bibinfo
  {author} {\bibfnamefont {S.}~\bibnamefont {Ogura}}, \ and\ \bibinfo {author}
  {\bibfnamefont {K.}~\bibnamefont {Fukutani}},\ }\href@noop {} {\bibinfo
  {journal} {submitted.}\ }\BibitemShut {NoStop}%
\bibitem [{\citenamefont {Diebold}(2003)}]{Diebold}%
  \BibitemOpen
\bibfield  {journal} {  }\bibfield  {author} {\bibinfo {author} {\bibfnamefont
  {U.}~\bibnamefont {Diebold}},\ }\href@noop {} {\bibfield  {journal} {\bibinfo
   {journal} {Surf. Sci. Rep.}\ }\textbf {\bibinfo {volume} {48}},\ \bibinfo
  {pages} {53} (\bibinfo {year} {2003})}\BibitemShut {NoStop}%
\bibitem [{\citenamefont {Zhang}\ \emph {et~al.}(2006)\citenamefont {Zhang},
  \citenamefont {Kawauchi}, \citenamefont {Fujimoto},\ and\ \citenamefont
  {Okano}}]{Zhang}%
  \BibitemOpen
  \bibfield  {author} {\bibinfo {author} {\bibfnamefont {X.}~\bibnamefont
  {Zhang}}, \bibinfo {author} {\bibfnamefont {T.}~\bibnamefont {Kawauchi}},
  \bibinfo {author} {\bibfnamefont {H.}~\bibnamefont {Fujimoto}}, \ and\
  \bibinfo {author} {\bibfnamefont {T.}~\bibnamefont {Okano}},\ }\href@noop {}
  {\bibfield  {journal} {\bibinfo  {journal} {Jpn. J. Appl. Phys.}\ }\textbf
  {\bibinfo {volume} {45}},\ \bibinfo {pages} {L142} (\bibinfo {year}
  {2006})}\BibitemShut {NoStop}%
\bibitem [{\citenamefont {Ruby}\ \emph {et~al.}(1963)\citenamefont {Ruby},
  \citenamefont {Hazoni},\ and\ \citenamefont {Pasternak}}]{Ruby}%
  \BibitemOpen
  \bibfield  {author} {\bibinfo {author} {\bibfnamefont {S.~L.}\ \bibnamefont
  {Ruby}}, \bibinfo {author} {\bibfnamefont {Y.}~\bibnamefont {Hazoni}}, \ and\
  \bibinfo {author} {\bibfnamefont {M.}~\bibnamefont {Pasternak}},\ }\href@noop
  {} {\bibfield  {journal} {\bibinfo  {journal} {Phys. Rev.}\ }\textbf
  {\bibinfo {volume} {129}},\ \bibinfo {pages} {826} (\bibinfo {year}
  {1963})}\BibitemShut {NoStop}%
\bibitem [{\citenamefont {Henke}\ \emph {et~al.}(1993)\citenamefont {Henke},
  \citenamefont {Gullikson},\ and\ \citenamefont {Davis}}]{Henke}%
  \BibitemOpen
  \bibfield  {author} {\bibinfo {author} {\bibfnamefont {B.~L.}\ \bibnamefont
  {Henke}}, \bibinfo {author} {\bibfnamefont {E.~M.}\ \bibnamefont
  {Gullikson}}, \ and\ \bibinfo {author} {\bibfnamefont {J.~C.}\ \bibnamefont
  {Davis}},\ }\href@noop {} {\bibfield  {journal} {\bibinfo  {journal} {At.
  Data. Nucl. Data Tables}\ }\textbf {\bibinfo {volume} {54}},\ \bibinfo
  {pages} {181} (\bibinfo {year} {1993})}\BibitemShut {NoStop}%
\bibitem [{\citenamefont {Head-Gordon.}\ \emph {et~al.}(1991)\citenamefont
  {Head-Gordon.}, \citenamefont {Tully}, \citenamefont {Schlichting},\ and\
  \citenamefont {Menzel}}]{Head}%
  \BibitemOpen
  \bibfield  {author} {\bibinfo {author} {\bibfnamefont {M.}~\bibnamefont
  {Head-Gordon.}}, \bibinfo {author} {\bibfnamefont {J.~C.}\ \bibnamefont
  {Tully}}, \bibinfo {author} {\bibfnamefont {H.}~\bibnamefont {Schlichting}},
  \ and\ \bibinfo {author} {\bibfnamefont {D.}~\bibnamefont {Menzel}},\
  }\href@noop {} {\bibfield  {journal} {\bibinfo  {journal} {J. Chem. Phys.}\
  }\textbf {\bibinfo {volume} {95}},\ \bibinfo {pages} {9266} (\bibinfo {year}
  {1991})}\BibitemShut {NoStop}%
\bibitem [{\citenamefont {Schlichting}\ and\ \citenamefont
  {Menzel}(1992)}]{SchlichtingSS}%
  \BibitemOpen
  \bibfield  {author} {\bibinfo {author} {\bibfnamefont {H.}~\bibnamefont
  {Schlichting}}\ and\ \bibinfo {author} {\bibfnamefont {D.}~\bibnamefont
  {Menzel}},\ }\href@noop {} {\bibfield  {journal} {\bibinfo  {journal} {Surf.
  Sci.}\ }\textbf {\bibinfo {volume} {272}},\ \bibinfo {pages} {27} (\bibinfo
  {year} {1992})}\BibitemShut {NoStop}%
\bibitem [{\citenamefont {Schlichting}\ \emph {et~al.}(1992)\citenamefont
  {Schlichting}, \citenamefont {Menzel}, \citenamefont {Brunner},\ and\
  \citenamefont {Brenig}}]{SchlichtingJCP}%
  \BibitemOpen
  \bibfield  {author} {\bibinfo {author} {\bibfnamefont {H.}~\bibnamefont
  {Schlichting}}, \bibinfo {author} {\bibfnamefont {D.}~\bibnamefont {Menzel}},
  \bibinfo {author} {\bibfnamefont {T.}~\bibnamefont {Brunner}}, \ and\
  \bibinfo {author} {\bibfnamefont {W.}~\bibnamefont {Brenig}},\ }\href@noop {}
  {\bibfield  {journal} {\bibinfo  {journal} {J. Chem. Phys.}\ }\textbf
  {\bibinfo {volume} {97}},\ \bibinfo {pages} {4453} (\bibinfo {year}
  {1992})}\BibitemShut {NoStop}%
\bibitem [{\citenamefont {Ikeda}(2013)}]{dt_ikeda}%
  \BibitemOpen
  \bibfield  {author} {\bibinfo {author} {\bibfnamefont {A.}~\bibnamefont
  {Ikeda}},\ }\href@noop {} {Ph.D. thesis},\ \bibinfo  {school} {The University
  of Tokyo} (\bibinfo {year} {2013})\BibitemShut {NoStop}%
\bibitem [{\citenamefont {Gr\"unsteudel}\ \emph {et~al.}(1999)\citenamefont
  {Gr\"unsteudel}, \citenamefont {Rusanov}, \citenamefont {Winkler},
  \citenamefont {Klaucke},\ and\ \citenamefont {Trautwein}}]{Grunsteudel}%
  \BibitemOpen
  \bibfield  {author} {\bibinfo {author} {\bibfnamefont {H.}~\bibnamefont
  {Gr\"unsteudel}}, \bibinfo {author} {\bibfnamefont {V.}~\bibnamefont
  {Rusanov}}, \bibinfo {author} {\bibfnamefont {H.}~\bibnamefont {Winkler}},
  \bibinfo {author} {\bibfnamefont {W.~M.}\ \bibnamefont {Klaucke}}, \ and\
  \bibinfo {author} {\bibfnamefont {A.~X.}\ \bibnamefont {Trautwein}},\
  }\href@noop {} {\bibfield  {journal} {\bibinfo  {journal} {Hyperfine
  Interact.}\ }\textbf {\bibinfo {volume} {122}},\ \bibinfo {pages} {345}
  (\bibinfo {year} {1999})}\BibitemShut {NoStop}%
\bibitem [{\citenamefont {Meixner}\ and\ \citenamefont
  {George}(1993)}]{Meixner}%
  \BibitemOpen
  \bibfield  {author} {\bibinfo {author} {\bibfnamefont {D.~L.}\ \bibnamefont
  {Meixner}}\ and\ \bibinfo {author} {\bibfnamefont {S.~M.}\ \bibnamefont
  {George}},\ }\href@noop {} {\bibfield  {journal} {\bibinfo  {journal} {Surf.
  Sci.}\ }\textbf {\bibinfo {volume} {297}},\ \bibinfo {pages} {27} (\bibinfo
  {year} {1993})}\BibitemShut {NoStop}%
\bibitem [{\citenamefont {Goldman}(1968)}]{Goldman}%
  \BibitemOpen
  \bibfield  {author} {\bibinfo {author} {\bibfnamefont {V.~V.}\ \bibnamefont
  {Goldman}},\ }\href@noop {} {\bibfield  {journal} {\bibinfo  {journal} {Phys.
  Rev.}\ }\textbf {\bibinfo {volume} {174}},\ \bibinfo {pages} {1041} (\bibinfo
  {year} {1968})}\BibitemShut {NoStop}%
\bibitem [{\citenamefont {Diebold}\ \emph {et~al.}(1998)\citenamefont
  {Diebold}, \citenamefont {Lehman}, \citenamefont {Mahmoud}, \citenamefont
  {Kuhn}, \citenamefont {Leonardelli}, \citenamefont {Hebenstreit},
  \citenamefont {Schmid},\ and\ \citenamefont {Varga}}]{DieboldSS}%
  \BibitemOpen
  \bibfield  {author} {\bibinfo {author} {\bibfnamefont {U.}~\bibnamefont
  {Diebold}}, \bibinfo {author} {\bibfnamefont {J.}~\bibnamefont {Lehman}},
  \bibinfo {author} {\bibfnamefont {T.}~\bibnamefont {Mahmoud}}, \bibinfo
  {author} {\bibfnamefont {M.}~\bibnamefont {Kuhn}}, \bibinfo {author}
  {\bibfnamefont {G.}~\bibnamefont {Leonardelli}}, \bibinfo {author}
  {\bibfnamefont {W.}~\bibnamefont {Hebenstreit}}, \bibinfo {author}
  {\bibfnamefont {M.}~\bibnamefont {Schmid}}, \ and\ \bibinfo {author}
  {\bibfnamefont {P.}~\bibnamefont {Varga}},\ }\href@noop {} {\bibfield
  {journal} {\bibinfo  {journal} {Surf. Sci.}\ }\textbf {\bibinfo {volume}
  {411}},\ \bibinfo {pages} {137} (\bibinfo {year} {1998})}\BibitemShut
  {NoStop}%
\end{thebibliography}
%merlin.mbs apsrev4-1.bst 2010-07-25 4.21a (PWD, AO, DPC) hacked
%Control: key (0)
%Control: author (8) initials jnrlst
%Control: editor formatted (1) identically to author
%Control: production of article title (-1) disabled
%Control: page (0) single
%Control: year (1) truncated
%Control: production of eprint (0) enabled
%

\end{document}